\documentclass[12pt,preprint]{aastex}

\usepackage{textcomp}
\usepackage{amsmath}
\usepackage{graphicx}
\usepackage{natbib}

\shorttitle{Thermal Stability of Magnetized Disks}
\shortauthors{Yu et al.}

\begin{document}

\title{Thermal Stability of Magnetized, Optically Thin, Radiative
Cooling-dominated Accretion Disks}

\author{Xiao-Fei Yu, Wei-Min Gu, Tong Liu, Ren-Yi Ma, and Ju-Fu Lu}

\affil{Department of Astronomy and Institute of Theoretical Physics
and Astrophysics, \\
Xiamen University, Xiamen, Fujian 361005, China}

\email{guwm@xmu.edu.cn}

\begin{abstract}
We investigate the thermal stability of optically thin, two-temperature,
radiative cooling-dominated accretion disks.
Our linear analysis shows that the disk is thermally unstable without
magnetic fields, which agrees with previous stability analysis
on the Shapiro-Lightman-Eardley disk.
By taking into account the effects of magnetic fields, however, we find that
the disk can be or partly be thermally stable.
Our results may be helpful to understand the outflows in optically thin flows.
Moreover, such radiative cooling-dominated disks may provide a new
explanation of the different behaviors between black hole and neutron
star X-ray binaries on the radio/X-ray correlation.
\end{abstract}

\keywords{accretion, accretion disks - black hole physics - instabilities
- magnetic fields - stars: winds, outflows}

\section{Introduction}

Accretion of rotating matter onto a compact object can provide a large amount
of released energy and is therefore believed to be the source of
cataclysmic variables, X-ray binaries, and active galactic nuclei (AGN).
The most famous accretion model is the geometrically thin and optically thick
disk (SSD) introduced by Shakura \& Sunyaev (1973).
Thermal instability was found in the inner region of SSD (Piran 1978)
and such an issue has been widely studied by linear analysis and
numerical simulations (Hirose et al. 2009; Lin et al. 2011, 2012;
Xue et al. 2011; Jiang et al. 2013; Zhu \& Narayan 2013).
For stellar-mass black holes, the temperature in such a standard disk
is in the range $10^4-10^7$~K,
which is quite low relative to the virial temperature.
In order to explain the hard X-ray spectra of black hole X-ray binaries
such as Cyg X-1, Shapiro, Lightman \& Eardley (1976) introduced an
optically thin, two-temperature disk model (hereafter SLE disk) with
$T_e \sim 10^9$K.
Unfortunately, shortly after the SLE disk was introduced, it was found that
the disk is thermally unstable (Pringle 1976; Piran 1978).
Both of the above two models are radiative cooling dominated, i.e., the
advective energy is negligible.

On the other hand, the energy released through dissipation may be trapped
within the accreted gas and then transported in the radial direction
towards the central object or stored in the flow as entropy.
Such accretion flows can be divided into two types, namely the slim disk
and the advection-dominated accretion flow (ADAF). The slim disk was
introduced by Abramowicz et al. (1988), which is optically thick with
extremely high mass accretion rates.
Recently, simulations on the slim disk have revealed that
strong outflows may occur (e.g., Ohsuga et al. 2005; Ohsuga \&
Mineshige 2011; Yang et al. 2014). Moreover, the radiative efficiency 
is not low and the luminosity can be far beyond the Eddington one
(Jiang et al. 2014).
The ADAF model was proposed by
Narayan \& Yi (1994), which is optically thin with low accretion rates.
The temperature of ADAF is close to the virial one, which is significantly
higher than that in the SLE disk due to energy advection.
Simulations on the ADAF have been studied by many previous works
(Narayan et al. 2012; Yuan et al. 2012a,b; for a review see
Yuan \& Narayan 2014).
Different from the stability of the SLE disk, both the slim disk
and the ADAF are thermally stable due to the dominant energy advection
(Abramowicz et al. 1995).
A recent work (Gu 2014), however, showed that the advective cooling
cannot balance the viscous heating, and therefore outflows ought to
be inevitable in such an accretion system.

The stability properties of accretion disks are of importance because
a global violently unstable disk may not exist in nature
and some instabilities restricted in a certain region of the disk
may contribute to the observed light variations in many systems.
Since the classic SLE disk was proved to be thermally unstable, such a type
of accretion disk is unlikely to be realized in nature.
It is generally believed that magnetic fields have fundamental influence on
the physics of accretion disks. For example, magnetorotational instability
(MRI) is known as a generator of turbulence, through which the angular momentum
can be transferred outwards (Balbus \& Hawley 1991).
Oda et al. (2009, 2010) presented a new thermal equilibrium solution
for optically thin disks by incorporating toroidal magnetic fields.
They argued that an optically thin, magnetic pressure-dominated accretion disk
(low-$\beta$ disk) will be thermally stable.
Zheng et al. (2011) revisited the thermal stability of standard thin disks
by including the role of toroidal magnetic fields.
Their calculation is based on the assumption of the perturbation relation
$\delta B_\varphi / B_\varphi = - \gamma \delta H / H$, where $\gamma$
is positive, i.e., the magnetic field will become weaker with increasing
height (or temperature).
This assumption is supported by the MHD simulation
of Machida et al. (2006) for a hot accretion flow, where they found that
the magnetic field becomes stronger when the disk shrinks vertically.
In addition, the large-scale magnetic fields in accretion disks have been widely
investigated and the strength may be underestimated particularly
for the region near the black hole (Cao 2011).

In this paper, we will analyze the thermal stability of optically thin disks
by including the role of magnetic fields.
The effects of such magnetic fields can be taken into account by
modifying the pressure as the sum of the gas pressure and the
magnetic pressure: $p=p_{\rm gas}+p_{\rm mag}$,
$p_{\rm mag}\sim B^2\sim \rho^{4/3}$ (Narayan \& Yi 1995; Yamasaki 1997).
We have to consider two-temperature plasma in such disks because
the electron temperature is expected to be significantly lower than
the ion temperature in such a low density, high temperature region.
In addition, the radiative cooling process is assumed to be
the thermal bremsstrahlung.
The paper is organized as follows.
The basic equations are presented in Section~2.
In Section~3, the thermal stability is investigated by linear analysis.
In Section~4, the thermal equilibrium curve is obtained by numerical
calculation.
Conclusions and discussion are made in Section~5.

\section{Equations}

We consider a steady state, optically thin, two-temperature black hole
accretion flows incorporating magnetic fields.
We assume the angular velocity is Keplerian, i.e., $\Omega=\Omega_{\rm K}$.
The disk structure is described by the following equations.
The continuity equation takes the form:
\begin{eqnarray}
&&\dot M = -4\pi RH\rho v_R~,
\end{eqnarray}
where $R$ is the cylindrical radius, $\rho$ the mass density of the accreted
gas, $H$ the half-thickness of the flow, $v_R$ the radial velocity,
which is defined to be negative when the flow is inward,
and $\dot{M}$ the mass accretion rate.
The equation of vertical hydrostatic equilibrium
can be written as
\begin{eqnarray}
&&H=\frac{1}{\Omega _{\rm K}} \left( \frac{p}{\rho} \right)^{1\over 2}~,
\end{eqnarray}
where $p$ is the pressure.
The Keplerian angular velocity $\Omega _{\rm K}$ takes the form
$\Omega _{\rm K}^2 = GM/(R-R_{\rm g})^2R$, where
the gravitational potential of the central black hole is assumed to be
$\Phi (R)=-GM/(R-R_{\rm g})$, which was introduced by
Paczy\'{n}ski \& Wiita (1980), with $M$ being the black hole mass
and $R_{\rm g}$ the gravitational radius, $R_{\rm g}$ $\equiv $ $2GM/c^2$.
From the angular momentum equation we can obtain the expression
of the radial velocity (e.g., Gu et al. 2006; Liu et al. 2007),
\begin{eqnarray}
&&v_R = \frac{\nu R^2}{\Omega R^2 - j}
\frac{d \Omega}{d R} \ ,
\end{eqnarray}
where $\nu$ is the kinetic viscosity coefficient,
and $j$ represents the specific angular momentum per unit mass accreted by
the black hole.

Since the radiation pressure is negligible in optically thin disks,
the total pressure $p$ is expressed as
\begin{eqnarray}
&&p=p_{\rm gas}+p_{\rm mag}~,
\end{eqnarray}
where $p_{\rm gas} = (1-\beta_{\rm mag})p$ is the gas pressure,
$p_{\rm mag} = \beta_{\rm mag}p$ the magnetic pressure.
The gas is assumed to consist of protons and electrons and therefore
the gas pressure $p_{\rm gas}$ is written as
\begin{eqnarray}
&&p_{\rm gas}=\frac{k_{\rm B}}{\mu m_{\rm p}}\rho (T_{\rm i}+T_{\rm e})~,
\end{eqnarray}
where $k_{\rm B}$ is the Boltzmann constant, $\mu$ the mean molecular
weight ($\mu=1$), $m_{\rm p}$ the proton mass, and $T_{\rm i}$,
$T_{\rm e}$ the ion and electron temperature, respectively.

To construct the energy equations, we introduce the following assumptions:
i) the dissipation energy goes into the ions;
ii) the energy is transferred from the ions to the electrons through
the Coulomb coupling;
iii) the electrons are cooled by the bremsstrahlung process.
Based on these assumptions, the energy equations of the ions and electrons
can be respectively written as
\begin{eqnarray}
&&q_{\rm vis}^+=\Lambda_{\rm ie}~,
\end{eqnarray}
\begin{eqnarray}
&&\Lambda_{\rm ie}=q_{\rm e}^-~,
\end{eqnarray}
where $q_{\rm vis}^+$ is the viscous heating rate per unit volume,
$\Lambda_{\rm ie}$ the energy transfer rate from the ions to the electrons
per unit volume, $q_{\rm e}^-$ the radiative cooling rate of electrons
per unit volume.
The viscous heating rate can be written as
\begin{eqnarray}
&&q_{\rm vis}^+=\rho \nu \left( R\frac{d \Omega}{d R}\right)^2.
\end{eqnarray}
We adopt the standard $\alpha$ prescription in this paper, i.e.,
$\nu=\alpha c_{\rm s} H$, where $c_{\rm s} \equiv (p/\rho)^{1/2}$ and
$\alpha$ is a constant parameter.
$\Lambda_{\rm ie}$ is given by Stepney \& Guilbert (1983):
\begin{eqnarray}
 \Lambda_{\rm ie} = \frac{3}{2} \frac{m_{\rm e}}{m_{\rm i}} n^2 \sigma_{\rm T} c  \left(\ln
  \Lambda\right) \frac{k T_{\rm i} - k  T_{\rm e}} {K_2(1/ \Theta_{\rm i})
  K_2(1/ \Theta_{\rm e})} \nonumber \\
  \left[ \frac{2(\Theta_{\rm i} + \Theta_{\rm e})^2
	      +1}{\Theta_{\rm i} + \Theta_{\rm e}} K_1
   \left( \frac{\Theta_{\rm i} + \Theta_{\rm e}}{\Theta_{\rm i}
    \Theta_{\rm e}} \right) + 2 K_0
   \left( \frac{\Theta_{\rm i} + \Theta_{\rm e}}{\Theta_{\rm i}
    \Theta_{\rm e}} \right)\right]~,
\end{eqnarray}
where $\sigma_{\rm T}$ is the Thomson scattering cross section and
$\ln \Lambda$ is the Coulomb logarithm (roughly $\ln \Lambda \sim 20$).
$K_n$ are modified Bessel function of the second kind of the order $n$,
respectively. The quantities $\Theta_{\rm i}$ and $\Theta_{\rm e}$
are defined as $\Theta_{\rm i}\equiv k_{\rm B}T_{\rm i}/m_{\rm p}c^2$
and $\Theta_{\rm e}\equiv k_{\rm B}T_{\rm e}/m_{\rm e}c^2$,
where $m_{\rm p}$ and $m_{\rm e}$ are the mass of the proton and electron,
respectively. For simplicity, we use the following formula
which uses no special functions (Kato et al.~2008):
\begin{eqnarray}
&&\Lambda_{\rm ie}=\frac{3}{2}\nu_{\rm E}\frac{\rho k_{\rm B}
(T_{\rm i}-T_{\rm e})}{m_{\rm p}}~,
\end{eqnarray}
with the electron-ion coupling being
$\nu_{\rm E}=2.4\times10^{21}(\ln \Lambda)\rho T_{\rm e}^{-3/2}$.

Following Narayan \& Yi (1995), the bremsstrahlung cooling rate
per unit volume is
\begin{eqnarray}
 q_{\rm e}^{-} = q_{\rm br}^{-} = q_{\rm br,ei}^{-} + q_{\rm br,ee}^{-}
= n^2 \sigma_{\rm T} c \alpha_{\rm f} m_{\rm e} c^2 \left[F_{\rm
  ei}(\Theta_{\rm e}) + F_{\rm
  ee}(\Theta_{\rm e}) \right]~,
\end{eqnarray}
where the subscripts ``$\rm ei$" and ``$\rm ee$" denote the electron-ion
and electron-electron bremsstrahlung cooling rates,
$\alpha_{\rm f}$ is fine-structure constant, and the function
$F_{\rm ei}(\Theta_{\rm e})$ and $F_{\rm ee}(\Theta_{\rm e})$
have the approximate form:
\begin{eqnarray}
 F_{\rm ei}(\Theta_{\rm e}) =
\left\{
\begin{array}{cc}
  \displaystyle \frac{9 \Theta_{\rm e}}{2 \pi} \left[ \ln ( 2 \eta
  \Theta_{\rm e} +
  0.48) + \frac{3}{2}\right] & (\Theta_{\rm e} > 1) \\ \\
  \displaystyle 4 \left( \frac{2 \Theta_{\rm e}}{\pi^3}\right)^{1/2} \left[ 1
  + 1.781 \Theta_{\rm e}^{1.34}\right] & (\Theta_{\rm e} < 1)
\end{array} \right. ~,
\end{eqnarray}
\begin{eqnarray}
 F_{\rm ee}(\Theta_{\rm e}) =
\left\{
\begin{array}{cc}
  \displaystyle \frac{9 \Theta_{\rm e}}{\pi} \left[ \ln ( 2 \eta \Theta_{\rm e}) +
  1.28\right] & (\Theta_{\rm e} > 1) \\ \\
  \displaystyle \frac{5}{6 \pi^{3/2}} (44 - 3 \pi^2 )\Theta_{\rm
   e}^{3/2} \times \nonumber \\
 ( 1
  + 1.1 \Theta_{\rm e} + \Theta_{\rm e}^2 - 1.25 \Theta_{\rm e}^{5/2}) &
  (\Theta_{\rm e} < 1)
\end{array}\right.  ~,
\end{eqnarray}
where $\eta = \exp(- \gamma_{\rm E}) $ and
$\gamma_{\rm E} \approx 0.5772$ is Euler's number.

\section{Thermal stability analysis}

The thermal instability criterion of accretion disks can be expressed as
(e.g., Frank et al. 1992)
\begin{eqnarray}
&&\left(\frac{\partial\ln Q^-}{\partial\ln T}\right)_{\rm \Sigma}<\left(\frac{\partial\ln Q^+}{\partial\ln T}\right)_{\rm \Sigma}~,
\end{eqnarray}
where $Q^+$, $Q^-$ and $T$ are respectively the viscous heating rate,
radiative cooling rate, and temperature.
$\Sigma$ is the surface density defined as
$\Sigma=2\rho H$. Such a criterion is identical to that in Piran (1978)
if we replace $T$ by $H$.

Since the disk is two-temperature, the thermal stability analysis should be
made for ion and electron temperature modulations, respectively.
The cooling timescale of the electron is quite short and therefore
the thermal stability of the electron is priority.
The surface density $\Sigma$ is assumed to be unchanged during a thermal
timescale, so we have
\begin{eqnarray}
&&\frac{d\ln \rho}{d\ln T_{\rm e}}=-\frac{d\ln H}{d\ln T_{\rm e}}~.
\end{eqnarray}
With Equations~(7), (10), (11) and (14), and the assumption that
$T_{\rm i}$ keeps unchanged during the perturbation of $T_{\rm e}$,
we can obtain
\begin{eqnarray}
&&\left(\frac{\partial\ln q_{\rm e}^+}{\partial\ln T_{\rm e}}\right)_{\Sigma}
- \left(\frac{\partial\ln q_{\rm e}^-}{\partial\ln T_{\rm e}}\right)_{\Sigma}
< 0~,
\end{eqnarray}
where $q_{\rm e}^+$ is equal to $\Lambda_{\rm ie}$.
The above relationship means that the disk is always stable against
electron temperature
perturbations. Such a result is in agreement with Kato et al.~(2008).

If electrons are thermally stable, then we have
\begin{eqnarray}
&&\frac{d(\ln q_{\rm e}^+)}{d\ln T_{\rm i}}=\frac{d(\ln q_{\rm e}^-)}{d\ln T_{\rm i}}~.
\end{eqnarray}
With Equation~(7) we can obtain the ratio of $T_{\rm i}$ to $T_{\rm e}$
as a function of $T_{\rm e}$, i.e., a function of $\Theta_{\rm e}$,
\begin{eqnarray}
&&\frac{T_{\rm i}}{T_{\rm e}} = g(\Theta_{\rm e})~.
\end{eqnarray}
Equation (2) gives
\begin{eqnarray}
&&\frac{d\ln p}{d\ln T_{\rm i}}=-\frac{d\ln \rho}{d\ln T_{\rm i}}~.
\end{eqnarray}
By using Equations~(4), (5) and (18), together with the relationship
$p_{\rm mag} \propto \rho^{4/3}$ mentioned in Section~1, we can derive
\begin{eqnarray}
&&\frac{d\ln \rho}{d\ln T_{\rm i}}=-\frac{3(1-\beta_{\rm mag})}
{6+\beta_{\rm mag}}\frac{d\ln (T_{\rm i}+T_{\rm e})}{d\ln T_{\rm i}}~,
\end{eqnarray}
then with Equations~(16), (17) and (19) we can obtain
\begin{eqnarray}
&&\frac{d T_{\rm e}}{d T_{\rm i}} = h(\beta_{\rm mag},\Theta_{\rm e})~.
\end{eqnarray}
Equations~(6), (8) and (10) can provide
\begin{eqnarray}
&&\frac{d\ln (q_{\rm i}^+- q_{\rm i}^-)}{d\ln T_{\rm i}}
= -3\frac{d\ln \rho}{d\ln T_{\rm i}}-\frac{d\ln (T_{\rm i}-T_{\rm e})}
{d\ln T_{\rm i}}+\frac{3}{2}\frac{d\ln T_{\rm e}}{d\ln T_{\rm i}}~.
\end{eqnarray}
Substituting Equations (17), (19) and (20) into Equation (21), we finally get
\begin{eqnarray}
&&\left[\frac{\partial \ln(q_{\rm i}^+-q_{\rm i}^-)}{\partial \ln T_{\rm i}}
\right]_{\rm \Sigma} = f(\beta_{\rm mag},\Theta_{\rm e})~,
\end{eqnarray}
then the thermal instability criterion can be expressed as
$f(\beta_{\rm mag}, \Theta_{\rm e})>0$.
Detailed expressions of $g(\Theta_{\rm e})$, $h(\beta_{\rm mag},
\Theta_{\rm e})$, and $f(\beta_{\rm mag}, \Theta_{\rm e})$
are presented in the Appendix.

Figure 1 shows the variation of the function $f(\Theta_{\rm e})$ for a
certain given value of $\beta_{\rm mag}$.
It is seen that for $\beta_{\rm mag}=0$, i.e., no magnetic field,
$f(\Theta_{\rm e})$ is positive for any $\Theta_{\rm e}$, which means that
the disk will be thermally unstable. This result agrees with
previous stability analysis on the SLE disk.
Actually, for $\beta_{\rm mag}=0$, the equations will reduce to
that for SLE disks.
The interesting result is that for $\beta_{\rm mag} = 0.7$,
$f(\Theta_{\rm e})$ is negative for any $\Theta_{\rm e}$, which implies
that the disk will be thermally stable. For the typical equipartition
case $\beta_{\rm mag} = 0.5$, the figure shows that there exists a thermally
stable region corresponding to $0.03 \la \Theta_{\rm e} \la 1$, whereas
for $\Theta_{\rm e} \la 0.03$ or $\Theta_{\rm e} \ga 1$ the disk will be
thermally unstable.
In other words, the disk can be or partly be thermally unstable
owing to the effects of magnetic fields.

The physical interpretation of thermally stable disks for moderate
strength of magnetic fields may be as follows. For a positive perturbation
of the ion temperature $T_{\rm i}$, the gas pressure $p_{\rm gas}$
will increase and therefore the disk thickness $H$ will also increase.
Consequently, the strength of magnetic fields $B$ and the magnetic pressure
$p_{\rm mag}$ will decrease due to the increasing volume.
Thus, the total pressure ($p_{\rm gas}+p_{\rm mag}$) as well as the
viscous stress will not increase as fast as in the disk without
magnetic fields. Therefore, the viscous heating rate $q_{\rm vis}^{+}$
may increase more slowly compared with the case without magnetic fields.
In this spirit, it is reasonable that the disk can become thermally 
stable when the role of magnetic fields is taken into account.

\section{Thermal equilibrium solutions}

In this section, we will show the numerical calculation of the local
thermal equilibrium solutions with Equations~(1)-(7), where
$M = 10 M_{\odot}$, $j = 1.8 cR_{\rm g}$, $\alpha=0.1$,
and $\beta_{\rm mag}=0.5$.
Figure~2 shows the radial variations of the ion and electron temperature
for $\dot m = 0.001$, $0.01$, and $0.1$,
where $\dot m \equiv \dot M/\dot M_{\rm Edd}$
($\dot M_{\rm Edd} \equiv 16 L_{\rm Edd}/c^2$).
It is seen that the ion temperature is significantly
higher than the electron temperature in the inner region of the disk.
We would like to point out that the temperature in the real case may be lower
than the current solutions since only the bremsstrahlung cooling process
is taken into consideration.
In the classic ADAF model where the radiative cooling is negligible
compared with the viscous heating, there exists a conservation
relationship (e.g., Molteni et al. 2001),
\begin{eqnarray}
&& \frac{1}{2} v_R^2 + \frac{\gamma}{\gamma-1} \frac{p}{\rho}
+ \frac{1}{2} \Omega^2 R^2 - \Omega (\Omega R^2-j)
- \frac{GM}{R-R_{\rm g}} = \lambda \ ,
\end{eqnarray}
where $\lambda$ is a constant and should be quite small compared with
the terms on the left hand side such as $GM/(R-R_{\rm g})$,
particularly for small radii.
Based on the above equation, we can derive the theoretical ion temperature
of ADAF by
\begin{eqnarray}
&& \frac{5}{2} \frac{k_{\rm B} T_{\rm i}}{\mu m_{\rm p}}
= \frac{GM}{R-R_{\rm g}} + \frac{1}{2} \Omega^2 R^2 - j\Omega \ ,
\end{eqnarray}
where $\gamma = 5/3$ and $\lambda = 0$ are adopted,
and the term $v_R^2/2$ is ignored. The theoretical ion temperature
is also plotted in Figure~2 by the navy blue line for a comparison. 
It is seen that the ion temperature of ADAF is significantly higher
than the cooling-dominated disks even for $\dot m = 0.1$, which indicates
that the energy advection is negligible and our thermal equilibrium solutions
are self-consistent.

Figure~3 shows the stable and unstable regions in the $\dot m-R$ diagram
for $\beta_{\rm mag} = 0.5$.
The region with $f(\Theta_{\rm e}) < 0$ corresponds to the
thermal stable solutions. For a given $\dot m$, the figure clearly shows
the thermally stable range in the disk.
It is seen that, for a typical accretion rate $\dot m = 0.01$, the disk
is thermally stable from the inner boundary to $\sim 10^3 R_{\rm g}$.
Such a stable solution may provide a second possibility for the optically
thin accretion flows.
For higher accretion rates such as $\dot m \ga 0.1$ the figure shows
an unstable inner region, which implies that outflows are likely to
be inevitable and the rate accreted by the black hole is less than
0.1. On the other hand, for low accretion rates such as $\dot m \la 0.001$,
the disk only has an inner stable part $\la 100 R_{\rm g}$. A thermally
stable disk for such low accretion rates may require stronger magnetic
fields, i.e., $\beta_{\rm mag} \to 0.7$. Another possibility for the outer
part of the disk to be stable may be related to the self-gravity, as
argued by Bertin \& Lodato (2001).

\section{Conclusions and Discussion}

In this work, we have investigated the thermal stability of magnetized,
optically thin, two-temperature, radiative cooling-dominated
accretion disks by linear analysis. We have derived a general criterion
for such an instability (Equation~22).
We have found that the disk is thermally unstable without magnetic fields,
which agrees with previous stability analysis on the SLE disk.
On the contrary, for adequately strong magnetic fields with
$\beta_{\rm mag}\sim 0.7$, the whole disk will be thermally stable.
For the typical equipartition case, $\beta_{\rm mag} = 0.5$,
the disk has a wide stable region particularly for $\dot m$ around $0.01$.
Our results may be helpful to understand the mechanism of outflows in
optically thin accretion flows, which may be triggered by the
thermal instability.

The well-known ADAF model is a classic model, and may be known as the unique
stable model for optically thin accretion flows. Since the black hole
has a horizon rather than a solid surface, it is easy to understand
that the flow will be transonic and the internal energy will
eventually be absorbed by the hole. On the contrary, for a neutron
star accretion system, since the neutron star has a real surface,
it remains uncertain whether the ADAF can still be stable and work
well in such a system, particularly for the inner region.
The present work provides a second possibility for the optically thin
accretion flows, i.e., the magnetized, radiative cooling-dominated
disk. In our opinion, since the radiative cooling can balance
the viscous heating locally in such a model, the energy advection
will be negligible. Thus, such a model may work well for both
black hole and neutron star systems.

The observations have shown that the radio and X-ray emission during the
hard state are strongly correlated for black hole and neutron star
X-ray binaries.
Taking the form of non-linear luminosity correlation, there exists
the relationship $L_{\rm R}\propto L_{\rm X}^b$,
where $L_{\rm R}$ is the radio luminosity and
$L_{\rm X}$ is the X-ray luminosity.
For neutron star X-ray binaries, the correlation index is $b \sim 1.4$
(Migliari \& Fender 2006).
On the contrary, for black hole X-ray binaries, the correlation
is quite complex.
At first the correlation with $b\sim 0.7$ was obtained
based on the data of different sources
(e.g., Hannikainen et al. 1998, Corbel et al. 2003, Gallo et al. 2003),
Later, however, another relation similar to that of neutron stars,
i.e., $b \sim 1.4$, was found for specific sources
(e.g., Coriat et al. 2011, Jonker et al. 2012, Ratti et al. 2012).
As discussed by Coriat et al. (2011), the steep correlation may indicate
that the disk is in the radiatively efficient regime.
In our point of view, the above observational results can be easily understood
based on our new model. Due to the existence of surface,
a neutron star system may prefer to the magnetized, radiative
cooling-dominated disk which corresponds to $b \sim 1.4$, whereas
a black hole system may have two choices, i.e., either the ADAF or the
cooling-dominated one, which corresponds to $b \sim 0.7$ and $\sim 1.4$,
respectively.
In addition, we would point out that, an alternative
cooling-dominated model for optically thin flows is the luminous hot
accretion flow (LHAF) introduced by Yuan (2001), which is, however, likely to be
thermally unstable. Concerning the stability, the magnetized, radiative
cooling-dominated disk studied in the present work is more likely
to exist than the other radiatively efficient disks.

\acknowledgments

We thank the referee for helpful comments that improved the paper.
This work was supported by the National Basic Research Program of China
(973 Program) under grant 2014CB845800, the National Natural Science
Foundation of China under grants 11103015, 11222328, 11233006, 11333004,
11473022, and U1331101,
the Fundamental Research Funds for the Central Universities
under grant 20720140532,
and the Natural Science Foundation of Fujian Province of China under
grant 2012J01026.

\appendix

\section{Expressions of thermal stability criterion}

In this Appendix, we will derive the detailed expressions of
$g(\Theta_{\rm e})$, $h(\beta_{\rm mag}, \Theta_{\rm e})$,
and $f(\beta_{\rm mag}, \Theta_{\rm e})$, which are mentioned in
Section~3.

\begin{eqnarray}
g(\Theta_{\rm e}) = 1 + \frac{\sigma_{\rm T}  \alpha_{\rm f} m_{\rm e}^{3/2}
m_{\rm p} c^4 \Theta_{\rm e}^{1/2} \left[F_{\rm ei}(\Theta_{\rm e})
+ F_{\rm ee}(\Theta_{\rm e}) \right]} {3.6\times10^{21} (m_{\rm p}
+ m_{\rm e})^2 k_{\rm B}^{3/2} \ln\Lambda} \ .
\end{eqnarray}

We define $X_1$, $X_2$, $Y_1$, $Y_2$, and $k$ as follows,
\[
X_1 = \Theta_{\rm e}\left[ \ln (2\eta\Theta_{\rm e}+0.48) + \frac{3}{2} \right]
+ 2 \Theta_{\rm e} \left[ \ln ( 2 \eta \Theta_{\rm e}) + 1.28\right] \ ,
\]
\[
X_2 = 8 (2 \Theta_{\rm e})^{1/2} (1 + 1.781\Theta_{\rm e}^{1.34})
+ \frac{5}{3} (44-3\pi^2)\Theta_{\rm e}^{3/2} (1 + 1.1 \Theta_{\rm e}
+ \Theta_{\rm e}^2 - 1.25 \Theta_{\rm e}^{5/2}) \ ,
\]
\[
Y_1 = \left[ \Theta_{\rm e}\ln (2\eta\Theta_{\rm e}+0.48)
+ 2\eta\Theta_{\rm e}^2/(2\eta\Theta_{\rm e}+0.48) + 6\Theta_{\rm e}
+ 2\Theta_{\rm e}\ln(2\eta\Theta_{\rm e})\right]/X_1 \ ,
\]
\[
Y_2 = \left[5.66\Theta_{\rm e}^{1.5} + 37.1\Theta_{\rm e}^{1.84}
+ 36\Theta_{\rm e}^{1.5} + 66\Theta_{\rm e}^{2.5}
+ 83\Theta_{\rm e}^{3.5} - 120\Theta_{\rm e}^4\right]/X_2 \ ,
\]
\[
k = \frac{3(1 - \beta_{\rm mag})}{6 + \beta_{\rm mag}} \ .
\]

We can derive the following expression:
\begin{eqnarray}
 h(\beta_{\rm mag}, \Theta_{\rm e}) =
\left\{
\begin{array}{cc}
  \displaystyle \frac {1/(g(\Theta_{\rm e})-1)-2k/(g(\Theta_{\rm e}) +1)}
  {1.5+Y_1+1/(g(\Theta_{\rm e})-1) + 2k/(g(\Theta_{\rm e}) +1)}
  & (\Theta_{\rm e} > 1) \\ \\
  \displaystyle \frac {1/(g(\Theta_{\rm e})-1)-2k/(g(\Theta_{\rm e}) +1)}
  {1.5+Y_2+1/(g(\Theta_{\rm e})-1) + 2k/(g(\Theta_{\rm e}) +1)}
  & (\Theta_{\rm e} < 1)
\end{array} \right. ~.
\end{eqnarray}

Finally, we obtain
\begin{equation}
\begin{split}
f(\beta_{\rm mag},\Theta_{\rm e})=\frac{3k g(\Theta_{\rm e})
(1+h(\beta_{\rm mag},\Theta_{\rm e}))}
{(g(\Theta_{\rm e})+1)} + 1.5 g(\Theta_{\rm e})
h(\beta_{\rm mag},\Theta_{\rm e})-\frac{g(\Theta_{\rm e})
(1- h(\beta_{\rm mag},\Theta_{\rm e}))} {(g(\Theta_{\rm e})-1)} \ ,
\end{split}
\end{equation}
which is the function introduced in Equation~(22).

\clearpage

\begin{figure}
\plotone{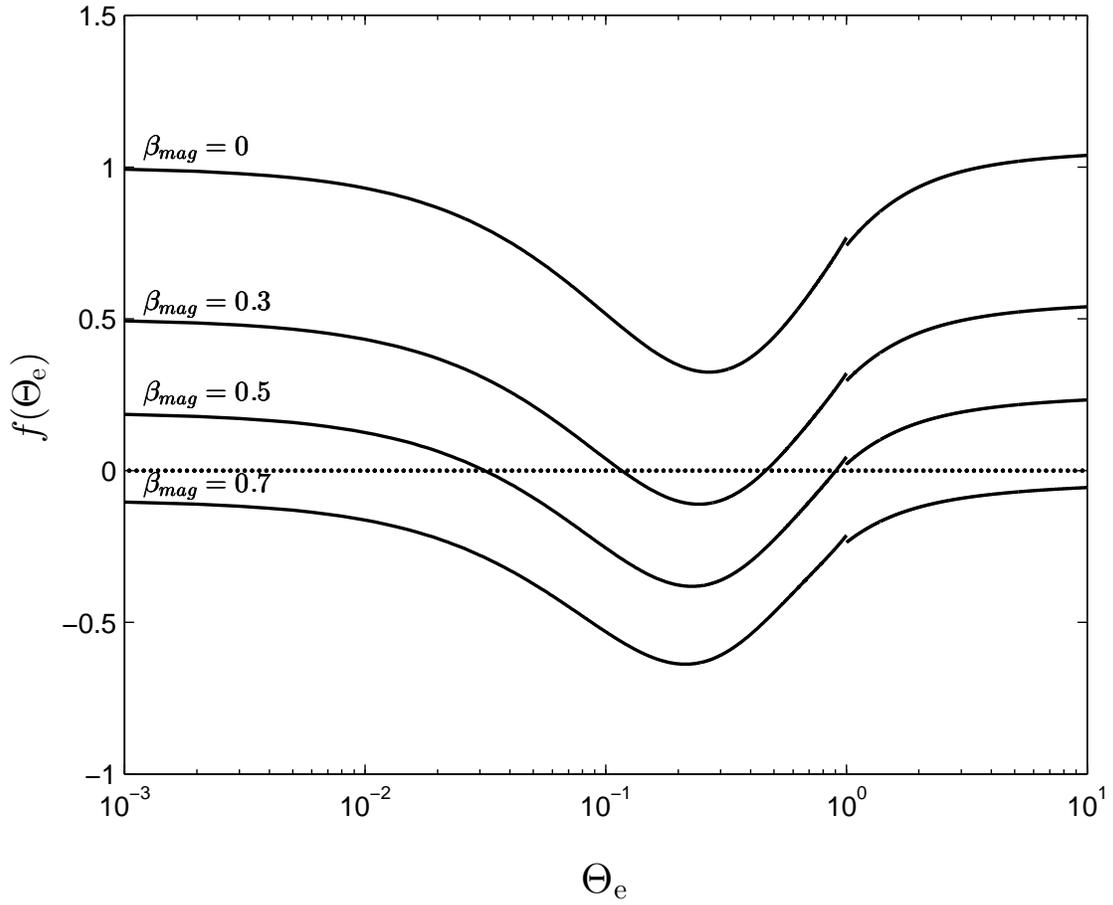}
\caption{
Variation of the function $f(\Theta_{\rm e})$ for $\beta_{\rm mag} = 0$,
0.3, 0.5, and 0.7.
}
\label{f1}
\end{figure}

\clearpage

\begin{figure}
\plotone{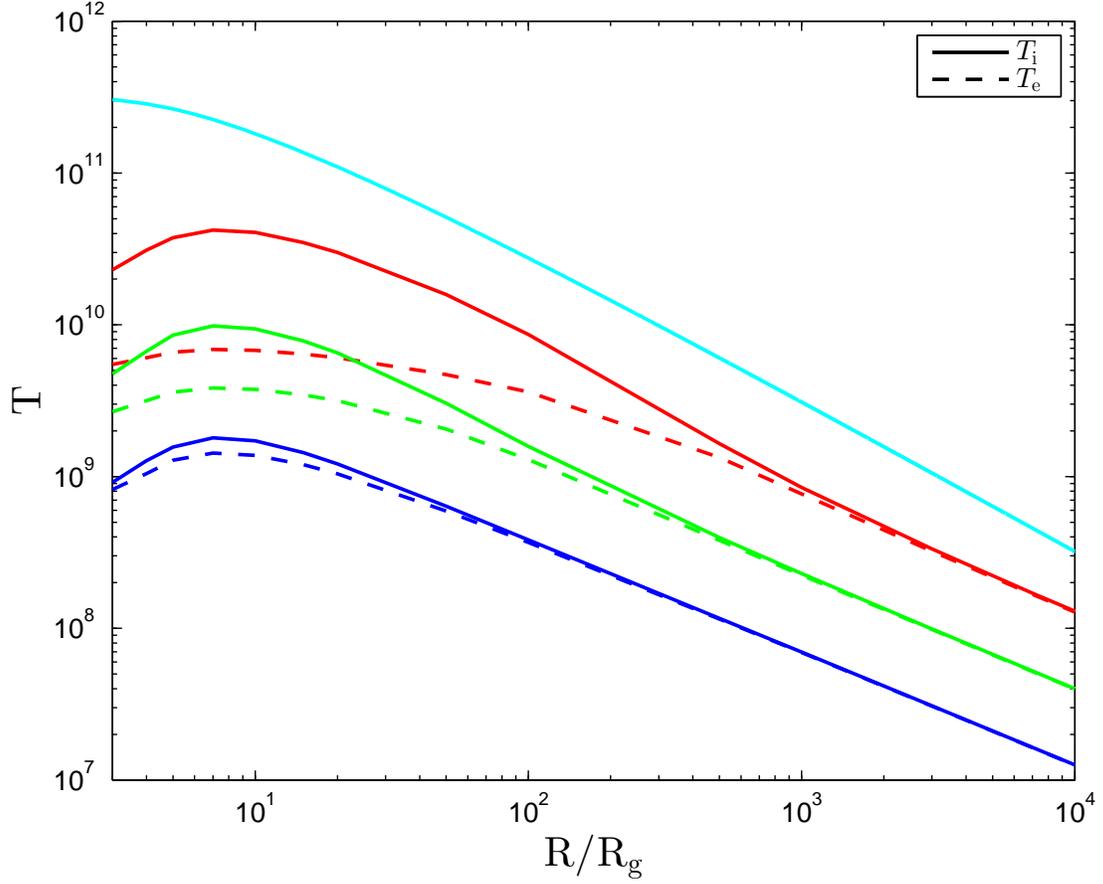}
\caption{
Radial variations of the ion temperature $T_{\rm i}$ (solid) and the
electron temperature $T_{\rm e}$ (dashed) for $\dot m = 0.001$ (blue),
$0.01$ (green), and $0.1$ (red), where $\beta_{\rm mag} = 0.5$.
The navy blue solid line represents the theoretical ion temperature
of ADAF (Equation~24).
}
\label{f2}
\end{figure}

\clearpage

\begin{figure}
\plotone{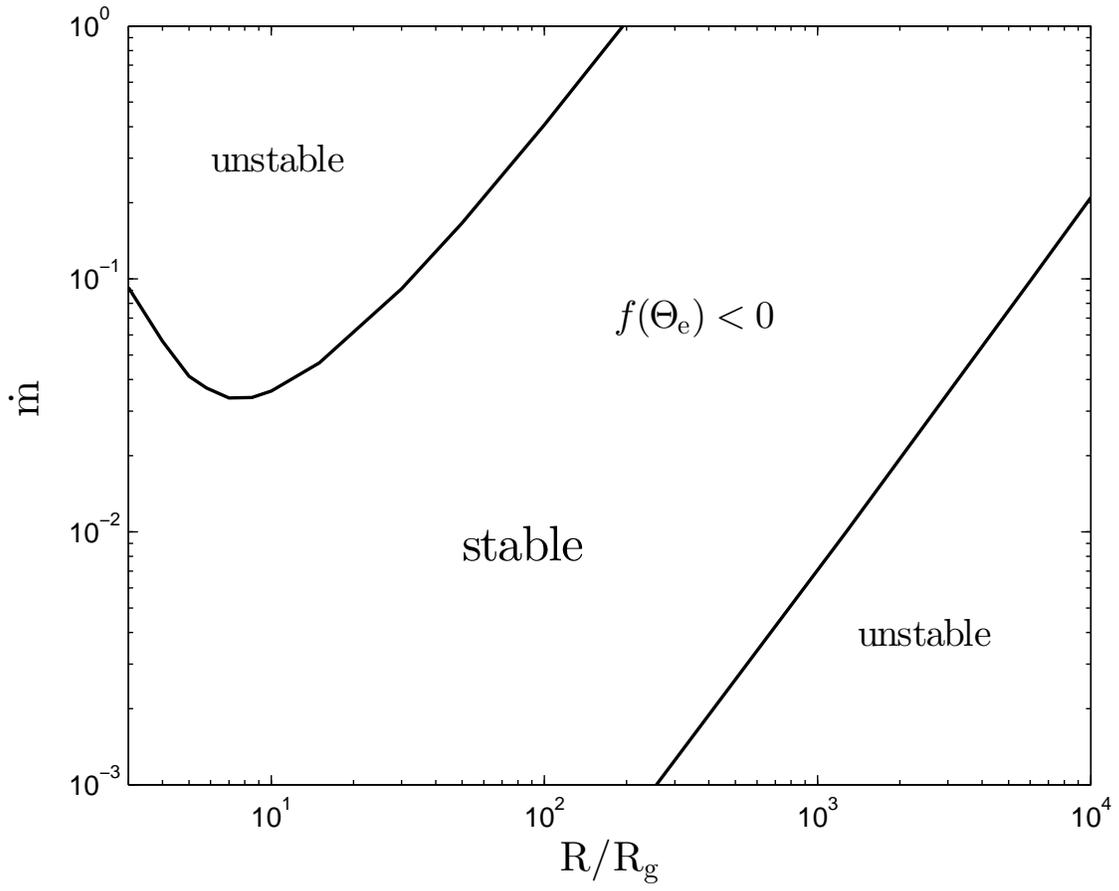}
\caption{
The $\dot m-R$ diagram for the thermal stability, where $\beta_{\rm mag}=0.5$.
}
\label{f3}
\end{figure}

\end{document}